# Coupled spin dynamics in epitaxial trilayer heterostructures of ferrimagnetic garnet

A. Del Giacco[1,2,3], M.J. Gross[4], O. Wojewoda[5,1], L. Menna[2,6], T. Grossmark[1], P. Lauer[1], V. Levati[2], S. Kurdi[6], E. Albisetti[2], D. Petti[2], M. Urbánek[5], and C.A. Ross[1*]

1. Department of Materials Science and Engineering, Massachusetts Institute of Technology, Cambridge, Massachusetts 02139, USA

2. Dipartimento di Fisica, Politecnico di Milano, Milano 20133, Italia

3. Institute for NanoSystems Innovation (NanoSI), Northeastern University, Boston, MA 02115, USA

4. Department of Electrical Engineering and Computer Science, Massachusetts Institute of Technology, Cambridge, Massachusetts 02139, USA

5. CEITEC BUT, Brno University of Technology, Brno, Czech Republic

6. Institute of Photonics and Quantum Sciences, SUPA, Heriot-Watt University, EdinburghEH14 4AS, United Kingdom

E-mail: caross@mit.edu

**The magnetization dynamics of an all-garnet trilayer consisting of $Y_3Fe_5O_{12}/Y_3Fe_3Al_2O_{12}/Y_3Fe_5O_{12}$ (YIG/YIAG/YIG) is analysed. The two magnetic YIG layers, separated by a 4 nm thick paramagnetic YIAG layer, are coupled via dipolar interactions leading to the formation of hybrid magnon modes distinct from the modes found in a single YIG layer. The YIAG exchange-decouples the YIG layers while enabling lattice coherence, maintaining the low damping of the YIG. Both ferromagnetic resonance and micro-Brillouin light scattering measurements were used to characterize the sample and its hybrid dynamics, which showed excellent agreement with an analytical reciprocal space model of spin-wave dynamics and with micromagnetic modeling of a dipolarly coupled magnetic heterostructure.**

## 1. Introduction

Magnetic heterostructures offer powerful pathways for engineering spin transport by tailoring interfacial exchange, spin–orbit interactions, and layer composition, enabling an expanding range of spintronic and magnonic functionalities. While metallic ferromagnets underpin many device concepts, insulating magnets, and in particular yttrium iron garnet ($Y_3Fe_5O_{12}$ or YIG) [1 - 6], have emerged as a uniquely capable platform, combining ultralow damping, a Curie temperature of 560 K, large optical transparency and high magneto-optical activity, and fast domain-wall dynamics [7]. These properties make YIG and other iron garnets (IGs) promising for low-power, high-speed magnonic, photonic, and radio-frequency technologies [8-14]. Heterostructures integrating IGs with metals have revealed a rich variety of interfacial phenomena, including spin Hall magnetoresistance, non-reciprocal spin-wave dynamics, spin-orbit-torque-driven domain wall motion, and magnon spin-valve effects [3, 15-23].

Despite this progress, epitaxial multilayers composed entirely of garnets remain poorly explored. Previous work has demonstrated garnet-based Bragg reflectors with layers tens of nm thick [24-26], superlattices with nm-thick layers, and bilayers incorporating rare-earth garnets [27-30]. We recently reported multilayers made of rare earth IGs with layers as thin as 0.5 nm, i.e. half a unit cell [31]. Interfacial Dzyaloshinskii-Moriya and exchange interactions have also been explored in garnet bilayers such as Bi-substituted YIG (BiYIG)/TmIG and YIG/TmIG with nm-thick layers [32, 33]. These works demonstrate the feasibility of making single crystal IG heterostructures, opening the possibility of using vertically stacked IG films with low damping for studies of magnon dynamics.

Coupling between magnon waveguides leads to hybridization of the excitations and can open a frequency gap for propagating magnons. Even in the absence of exchange coupling, stray field interactions can produce dipolar-coupled modes. Significant attention has been devoted to coupled spin dynamics in in-plane magnonic architectures such as closely spaced pairs of lithographically-defined waveguides [34-38]. These structures exhibit dynamic dipolar coupling that hybridizes the fundamental modes into symmetric and antisymmetric branches, an effect exploited in magnonic crystals, directional couplers, and reconfigurable spin-wave circuitry. Such in-plane systems have established the canonical framework for describing dipolar-mediated mode splitting, bandgap formation, and spin-wave localization.

Analogous effects are also expected in stacked magnetic waveguides separated by a nonmagnetic spacer. Theoretical studies predict rich coupled-mode phenomenology in such vertical trilayers [39-43], but experimental investigations to date have been restricted to metallic trilayers [44-52] or heterostructures of IGs with other materials as the spacer. For example, magnon transport has been studied in metallic layers of CoFeB exchange-coupled via Ru or W to form synthetic antiferromagnets [51-56], showing the formation of optical and acoustic hybrid modes. In exchange-coupled bilayers such as BiYIG/NiFe the gap opening varied with field orientation [21], and non-linear and non-reciprocal effects have been reported in thick bilayers of substituted YIG [57,58]. Magnon transport and spin-valve effects have been reported in YIG heterostructures including YIG/NiO/YIG, YIG/CoO/Co, YIG/Au/YIG and YIG/Pt/YIG [59-65]. In TmIG/MgO/TmIG bilayers with perpendicular magnetic anisotropy, a magnon frequency gap was observed when the in-plane field was orthogonal to the magnon wave vector indicative of a strong coupling regime [66]. Multilayers made of YIG are desirable to minimize damping and to access the intrinsic coupled-magnon regime [67-69], but the insertion of metallic or non-garnet spacers increases damping and disrupts the epitaxial growth of the upper layer of YIG, limiting the performance. This motivates the development of fully epitaxial YIG heterostructures where the spacer layer is a nonmagnetic garnet for studies of magnon propagation and hybridization.

Here, we introduce an epitaxial all-garnet heterostructure that enables access to coupled-magnon dynamics. The vertically engineered garnet heterostructures are composed of two high-quality epitaxial YIG films dipolar-coupled through a few-nanometer-thick nonmagnetic garnet spacer. Such a system is expected to host two distinct, higher and lower frequency, branches, characterized by symmetric and anti-symmetric oscillating modes of the magnetization in each layer, according to the spin wave configuration. In particular, at a fixed frequency, the wavevector separation $\Delta k_x = k_s - k_{as}$ where $k_s$ and $k_{as}$ represent wavevectors of the symmetric and antisymmetric modes, determines interference between the modes that mediates the periodic exchange of energy between the two magnetic layers, which occurs over a distance $\approx \pi/\Delta k_x$[37,43].

The vertically engineered garnet heterostructures in this work consist of $Y_3Fe_5O_{12}$ (95 nm) / $Y_3Fe_3Al_2O_{12}$ (4 nm) / $Y_3Fe_5O_{12}$ (93 nm) grown on a $Gd_3Ga_5O_{12}$ (111) substrate, in which the $Y_3Fe_3Al_2O_{12}$ (YIAG) spacer exchange-decouples the two YIG layers while preserving their magnetostatic coupling and structural quality. By adjusting the aluminum content, the Curie temperature and the magnetic response of the interlayer can be tuned and its lattice matching to YIG optimized. Moreover, the interlayer introduces a non-negligible interfacial anisotropy which can be engineered to break structural symmetry and induce non-reciprocal phenomena. This new class of garnet-based magnonic heterostructures therefore provides a platform for exploring non-reciprocal, coupled spin-wave dynamics in insulating, low-damping multilayers.

## 2. Results and Discussion

### 2.1. Growth and Characterization of the $Y_3Fe_5O_{12}$ / $Y_3Fe_3Al_2O_{12}$/ $Y_3Fe_5O_{12}$ Heterostructure

A central challenge of thin film YIG growth is to ensure lattice matching required for high-quality epitaxial growth to achieve low magnetic damping. YIG (cubic, lattice parameter $a_o$(YIG) = 1.2376 nm) is well matched to gadolinium gallium garnet (GGG, $a_o$(GGG) = 1.2383 nm). GGG is a possible candidate for the spacer layer in a YIG heterostructure, but we find that it tends to grow with a rough surface under the deposition conditions used for YIG, and interdiffusion affects the properties of the YIG. YAG ($Y_3Al_5O_{12}$) is nonmagnetic but its lattice parameter is $a_0 = 1.201$ nm, substantially smaller than YIG and GGG, leading to a strained and defective layer. Instead, we employ co-deposition from YIG and YAG to form a solid solution YIAG ($Y_3(Fe,Al)_5O_{12}$) as a non-magnetic spacer separating the undoped YIG layers. We select a YIG:YAG composition with sufficient Al to reduce the Curie temperature $T_C$ below room temperature while maintaining the lattice parameter to be as close to that of YIG as possible. A molecular field coefficient model [3] was used to determine Curie temperature and magnetic response as a function of Al content. A ratio of 3:2 Fe:Al yields $T_C$ = 294 K and a negligible room temperature magnetic moment (see Supporting InformationS1), with a lattice parameter of $a_0 = 1.223$ nm.

The resultant heterostructure consists of $Y_3Fe_3O_{12}$/$Y_3Fe_3Al_2O_{12}$/$Y_3Fe_3O_{12}$ grown via PLD from stoichiometric YIG and YAG targets onto (111)-oriented $Gd_3Ga_5O_{12}$ (GGG) substrates (Experimental Methods). Figure 1 presents the structural and compositional characterization of the YIG/YIAG/YIG heterostructure. High-resolution X-ray diffraction of the (444) reflection (Fig. 1a) reveals pronounced Laue fringes surrounding the substrate peak, consistent with a uniform film thickness and smooth interfaces of the trilayer stack. The diffraction peaks from the two YIG layers coincide with the substrate peak. Fitting of the trilayer–substrate diffraction pattern yields YIG layer thicknesses of 95 nm and 93 nm, and spacer YIAG layer thickness of 4 nm, in excellent agreement with the thicknesses expected from the measured growth rates.

High angle annular dark field (HAADF) scanning transmission electron microscopy (STEM) images (Fig. 1c,d) demonstrate the sharpness of the YIG/YIAG interfaces and the preservation of epitaxy across the heterostructure. The composition contrast between the YIG layers and the YIAG spacer layer is clearly visible in the HAADF image corresponding to the Al and Fe rich regions seen in the energy-dispersive X-ray spectroscopy (EDX) mapping, (Fig. 1e), indicating negligible Al interdiffusion into the adjacent YIG layers. The layer thicknesses are in agreement with those derived from fits to the X-ray diffraction data.

The magnetic hysteresis loop of the trilayer heterostructure, measured with the external magnetic field applied in the film plane, is shown in Fig. 1b. The film has an in-plane easy axis with a saturation magnetization of $M_s = 144 \pm 5$ kA/m and a coercive field $\mu_0 H_c < 0.1$mT, consistent with high-quality YIG films.

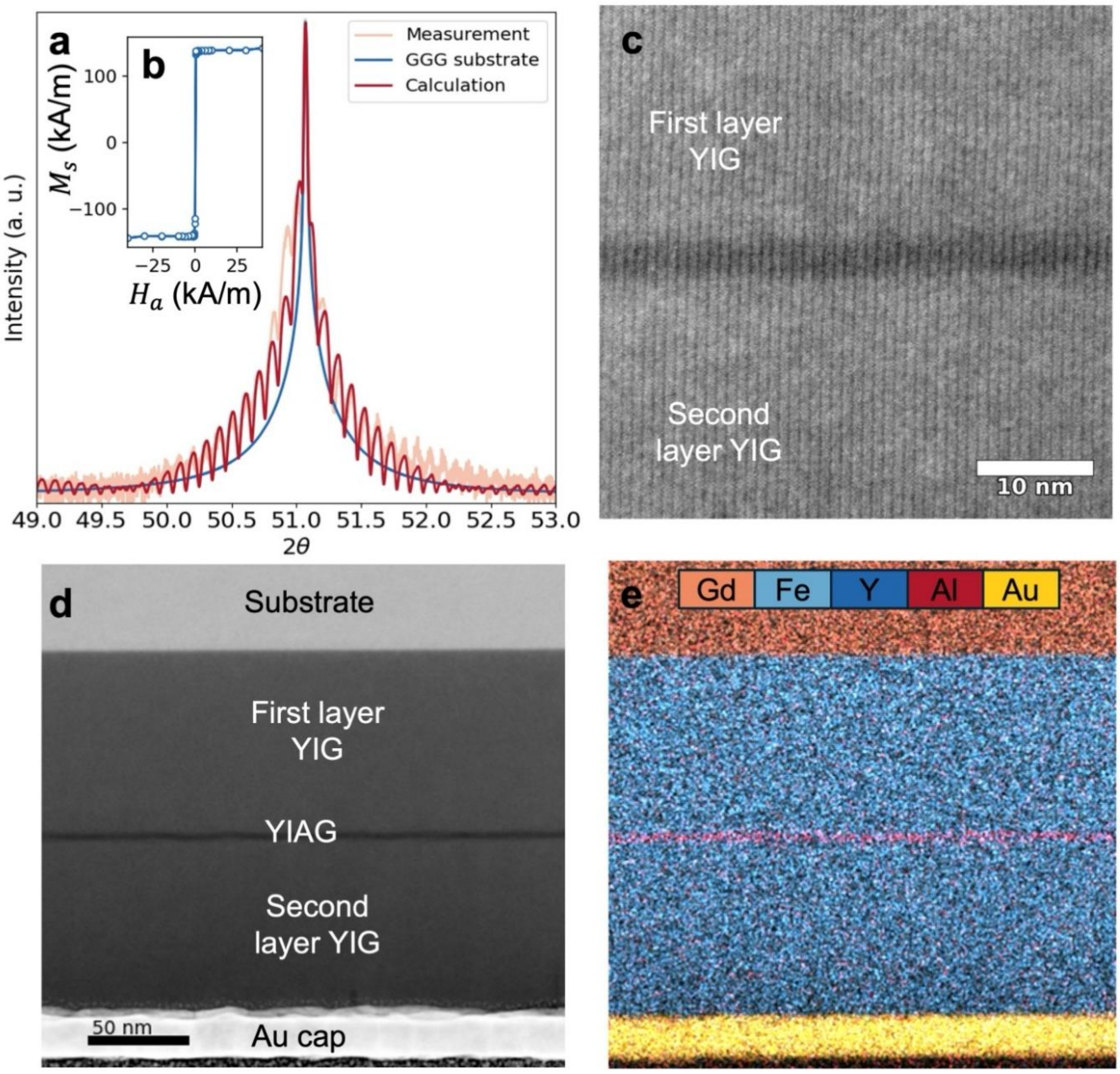

**Figure 1.** ((Structural and magnetic characterization of the YIG/YIAG/YIG trilayer.))

**a** High-resolution X-ray diffraction (XRD) scan around the (444) reflection of the GGG substrate (blue line). The well-defined Laue oscillations (red line: calculation; orange line: measurement) confirm the high crystalline quality and sharp interfaces of the stack.

**b** In-plane magnetization hysteresis loop of the heterostructure measured by vibrating sample magnetometry (VSM), exhibiting a square loop with saturation magnetization consistent with bulk YIG.

**c** High-resolution transmission electron microscopy (HR-TEM) image including the thin YIAG spacer layer, demonstrating epitaxial growth across the interfaces.

**d** Cross-sectional scanning transmission electron microscopy (STEM) overview and **e** corresponding energy-dispersive X-ray spectroscopy (EDX) elemental mapping, showing the distribution of Al (red), Fe (light blue), Y (dark blue), Gd (orange), and Au (yellow).

## 2.2. Heterostructure Modeling

To model the system dynamic we exploit both analytical spin-wave theory and micromagnetic simulations, extending the analytical framework established for in-plane coupled waveguides [37] to describe the YIG/YIAG/YIG heterostructure. The analytical model is summarized here and described in detail in [70]. Micromagnetic simulations are conducted using Ubermag and TetraX [71-73] micromagnetic codes (see Experimental Methods).

Starting from the linearization of the Landau–Lifshitz equation it is possible to rewrite the problem in reciprocal space, incorporating exchange, self- and mutual dipolar interactions (vanishing at $k \approx 0$ for an infinite film), Zeeman energy and the uniaxial anisotropy terms:

$$-j\omega_{k_x}^{p}\,\hat{\mathbf{I}}\vec{m}_{k_x}^{p} = \vec{\mu} \times \sum_q \widehat{\boldsymbol{\Omega}}_{k_x}^{p,q}\,\vec{m}_{k_x}^{q} \tag{2}$$

where, $p, q \in \{1,2\}$ index the magnetic layers, $\hat{\mathbf{I}}$ is the identity matrix, and $\widehat{\boldsymbol{\Omega}}_{k_x}^{p,q}$ is the dynamical tensor operator collecting the aforementioned terms [74-76]. Solving the associated eigenvalue problem yields both the eigenfrequencies and the eigenvector of the system, allowing the reconstruction of the SW dispersion and associated spatial profiles of the dynamic magnetization. We exploit this framework to quantitatively assess the agreement between the coupled-mode description and the experimental observations.

The inequivalent interfaces of the two magnetic layers (air and YIAG for the top layer vs. YIAG and GGG substrate for the bottom layer) lead to a difference in their magnetic properties. This interfacial anisotropy difference is treated in the modeling by introducing a uniaxial out-of-plane term, $K_u$, present only in the top layer [77]. This anisotropy difference is expected to manifest as distinct FMR resonances from the two YIG layers, and as a gap opening between the symmetric and antisymmetric branches for finite wavevectors. To quantify the effect, we calculated the FMR frequencies vs. applied magnetic field, as shown in Figure 2a for $K_u$ = 0 – 10 kJ m$^{-3}$. The resonances can be attributed to the bottom ($K_u$= 0) and top ($K_u \neq 0$ ) layers (See Supporting information S2).

Extending this calculation to finite wave vectors allows us to associate the mode profiles with either the symmetric or antisymmetric branches. Inspection of Figure 2b,c reveals a notable feature: the inversion of the collective character of the dynamic magnetization between the Damon-Eshbach (DE) and Backward Volume (BV) configurations. In the DE configuration, the symmetric and antisymmetric modes correspond to the higher and lower frequency branches, respectively, whereas in the BV configuration, the opposite. Consequently, at a fixed excitation frequency, the symmetric mode consistently corresponds to the smaller wave vector and, therefore, the longer wavelength.

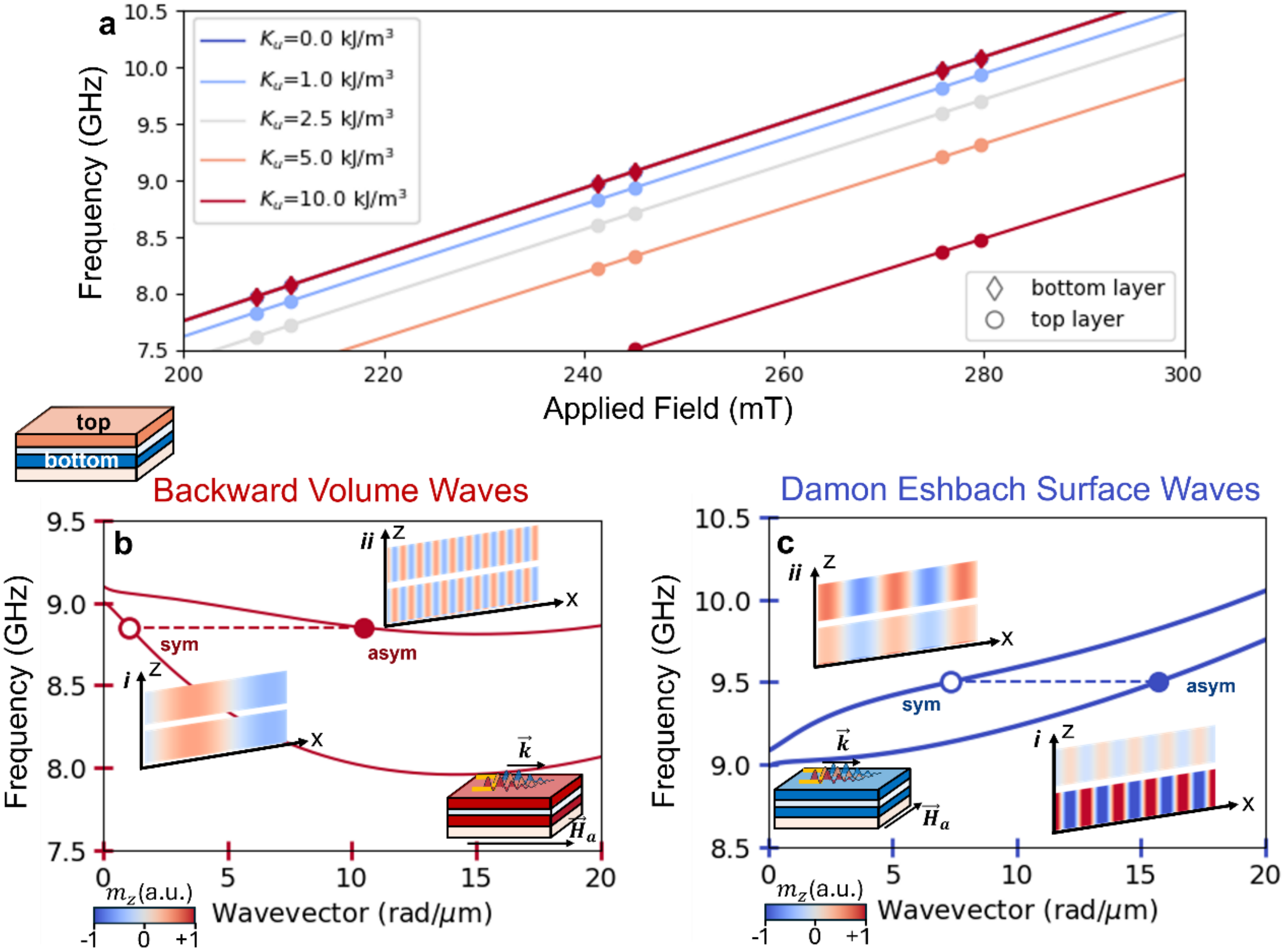

**Figure 2.** ((Dynamical modeling))

**a**, Simulated FMR response of the system in function of the anisotropy of the top layer at the experimental applied magnetic fields. **b–c**, Dispersion relations in the positive-wavevector region in DE and BV configurations for an applied field $H_a = 245$ mT. In insets (***i, ii***), two finite non-zero wavevectors from the symmetric (empty dot) and antisymmetric (full dot) branches were selected at 8.85 GHz in the BV configuration and 9.5 GHz in the DE configuration, respectively. These were chosen to highlight the symmetric and antisymmetric character of the oscillating out-of-plane magnetization component ($m_z$) at the resonant frequency for each spin-wave configuration, normalized between the two magnetic layers (see methods).

### 2.3. Spin-Dynamic Properties of the Heterostructure

Ferromagnetic resonance (FMR, Figure 3) and micro-focused Brillouin light scattering ($\mu$BLS, Figure 4) measurements were used to probe the dynamical behavior of the YIG/YIAG/YIG. Zero-wavevector ($k = 0$) analysis was done via field-frequency derivative FMR (see Experimental Methods). Thermally excited spin-wave spectroscopy (thermal BLS), in the perpendicular polarization scheme, was used to access the overall spectral landscape and identify the fundamental and higher-order modes associated with the heterostructure dynamics. Then phase-resolved $\mu$BLS measurements were carried out at finite-wavevector ($k \neq 0$), using sub-micron strip line RF transducers directly deposited on top of the trilayer.

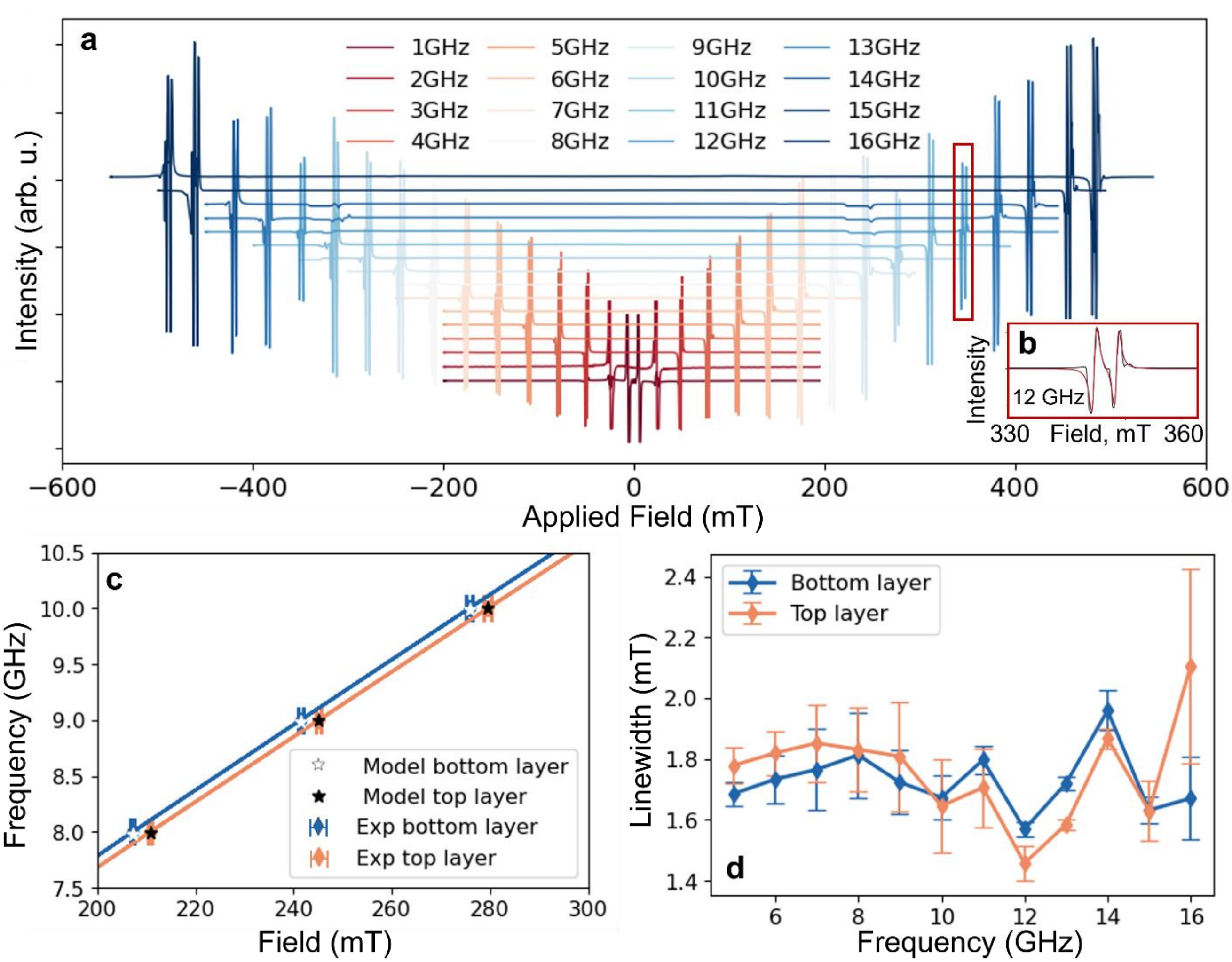

**Figure 3** (( FMR characterization ))

**a,** Broadband FMR spectra acquired by sweeping the excitation frequency from 1 to 16 GHz as a function of the applied magnetic field. Traces are vertically offset for clarity. **b,** Inset of the response measured at 12 GHz as a function of the applied external magnetic field, together with the corresponding fit, revealing the presence of two distinct resonance peaks named as optical and acoustic branches.

**c,** Top and bottom layer resonance fields extracted from the FMR data in Fig. 3a and plotted as a function of the external magnetic field for each excitation frequency, with the associated simulated data (stars) superimposed. At fixed frequency, increasing the magnetic field shifts the dispersion relations upward, bringing first the bottom ($K_u$= 0 kJ/$m^3$) and subsequently the top layer ($K_u = 0.65$ kJ/$m^3$) into resonance.

**d,** Extracted linewidths of the two layers resonances of the data shown in panel **b**.

### 2.3.1 Ferromagnetic Resonance

Figure 3a reports the FMR data between 1 and 16 GHz for a field sweep between $\pm 400$ mT. The measurements were carried out exploiting anharmonic sampling, ensuring an external field resolution of 0.1 mT in the resonance regions (see Experimental Methods). A magnified view of the spectrum at 12 GHz is presented in Figure 3b, where two distinct resonances are clearly resolved between 340 and 350 mT.

To quantitatively track the resonant fields (Figure 3c) and the associated linewidths, the spectra were fitted by a Lorentzian derivative consisting of two superposed lines with amplitude $A_1$ (left resonance) and $A_2$ (right resonance):

$$f(H) = A_1 \frac{(H-H_1)\Delta H_1}{((H-H_1)^2+(\Delta H_1/2)^2)^2} + A_1 \frac{(H-H_2)\Delta H_2}{((H-H_2)^2+(\Delta H_2/2)^2)^2}\,, \qquad (1)$$

where $H_1$and $H_2$ are the left and right resonant fields and $\Delta H_1, \Delta H_2$ their full width at half maximum, yielding a linewidth DH of approximately 1.8 mT for both peaks, Figure 3d. This may be compared with values of 0.5 mT or more reported for YIG films[78].

We remark that there is no dipolar coupling between these zero-wavevector modes. Instead, the observation of two distinct resonance peaks is a result of a difference in properties such as anisotropy between the two layers which lifts the degeneracy of the uniform mode, in accordance with the calculation in Figure 2a. The agreement between the experimental data and the analytical model evaluated for $K_u$= 0.65 kJ m$^{-3}$ in the top layer is shown in Fig. 3c.

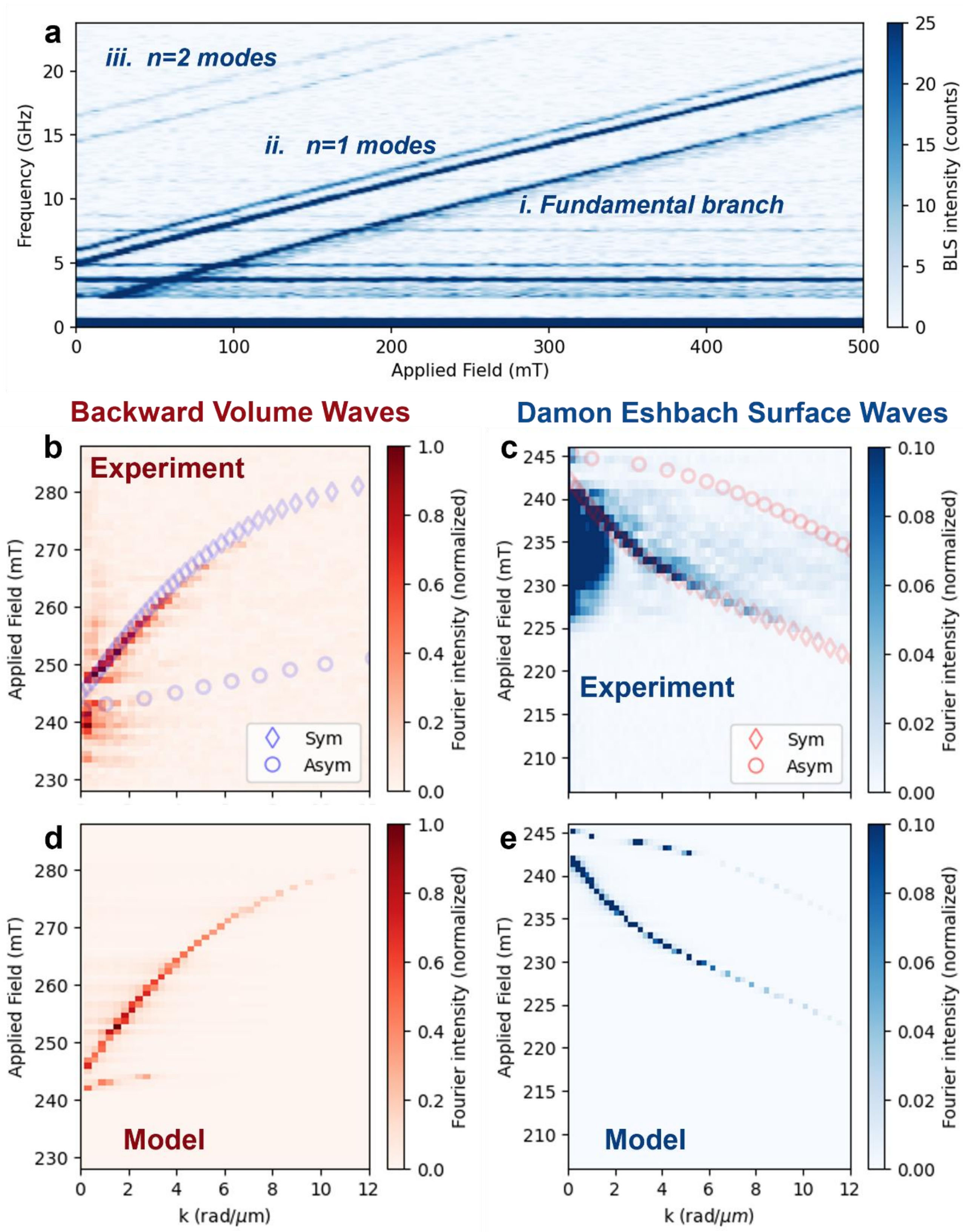

**Figure 4.** (($\mu$BLS characterization ))

**a,** Thermal Brillouin light scattering (BLS) spectra measured while sweeping the external magnetic field from 0 to 500 mT, showing the fundamental modes within the broadened region (i) and higher-order modes split into pairs, due to the presence of two YIG layers (ii, iii).

**b,c**, Phase-resolved BLS measurements acquired at an excitation frequency of 9 GHz in the backward volume and Damon Eshbach configurations, respectively. The data are obtained via spatial scans performed with a piezoelectric stage step of 250 nm, yielding a wavevector resolution of 0.3 rad/μm. To facilitate comparison with the analytical model of **d,e**, two overlays of the simulated data corresponding to the frequencies of the two expected branches are provided.

**d,e**, Simulated phase maps of the dynamic magnetization derived from the eigenvector analysis at each magnetic field for a fixed excitation frequency of 9 GHz. They are obtained by Fourier transforming the spatial magnetization profiles and accounting for the RF efficiency weighting induced by the finite excitation window of the transducer, in agreement with the experimental attenuation observed in **b,c.**

### 2.2.2 Thermal BLS

In thermal $\mu$BLS, spin waves are excited over a broad wavevector range, while the detected wavevector is determined by the optical scattering geometry, especially the numerical aperture (NA) of the objective lens. In our system (NA=0.75) the signal is predominantly formed by spin waves in the range of 0-12 rad/μm [79]. Figure 4a presents the thermal $\mu$BLS spectra upon external magnetic field sweep between 0 and 500 mT. All the observed magnetic mode frequencies increase monotonically with the applied field, indicating the absence of any field-induced magnetization reorientation. In addition to the fundamental branch (see label *i* in Figure 4a), higher-order modes are clearly resolved and appear as pairs (see labels *ii, iii* Figure 4a). This behavior is consistent with the slightly different magnetic properties of the two YIG layers seen in the FMR data or the different thicknesses, giving rise to two distinct sets of perpendicular standing spin-wave (PSSW) modes. The separation between corresponding modes increases with mode order, and is unresolved for the fundamental branch.

In addition to the magnetic field-dependent modes, field-independent features are present in the spectra, attributed to phononic contributions and instrumental background. Owing to the finite experimental resolutions of approximately 125 MHz in frequency and external magnetic field step of 5.5 mT, the thermal $\mu$BLS measurements do not resolve the small frequency and field separation expected between the fundamental symmetric and antisymmetric modes of a coupled magnetic heterostructure, nor allow a quantitative determination of the intrinsic magnetic damping from thermal $\mu$BLS linewidths.

### 2.2.3. Phase Resolved $\mu$BLS

To investigate the dynamics of coupled modes we fabricated microwave transducers based on strip-lines with conductor widths of 100 nm and 200 nm (see Experimental Methods) and conducted single-side phase-resolved BLS measurements. The excitation frequency was fixed at 9 GHz, while the external magnetic field was swept in steps of 1 mT between 228 and 288 mT in backward-volume (BV) configuration and in steps of 0.7 mT between 206 and 246 mT in the Damon–Eshbach (DE) configuration. To account for a 5% discrepancy in the applied magnetic field calibration between the BLS and FMR setups, we applied a shift of approximately 5 mT to the BLS data to align it with the FMR measurements and the simulations.

The spin-wave signal measured by phase-resolved μ-BLS was Fourier transformed to obtain field–wavevector ($H$–$k$) maps of the propagating dynamics (Fig. 4b,c). Spatial scans were performed using a piezoelectric stage over distances of 20 $\mu$m with a step size of 250 nm, corresponding to a wavevector resolution of approximately 0.3 rad/μm (see Supporting Information S4).

While the symmetric modes are always visible (within the antenna efficiency), such as the two distinct FMR peaks, the antisymmetric modes are observable only at low wavevectors. This behavior can be understood as a consequence of the spatial homogeneity of the excitation field across the out-of-plane direction, which prevents the excitation of antisymmetric modes. Indeed, the experimental observation of the two distinct FMR modes, and the detection of antisymmetric modes only at small k, is consistent with the lack of a defined symmetry of the dynamic magnetization in this spectral region. In this regime the layers experience a vanishing dipolar coupling ($k \approx 0$), where a precise phase relationship between the layers has not yet fully developed. Consequently, our experiment is capable of retrieving all modes for which the excitation field exerts a non-zero net torque, i.e. those lacking a well-defined symmetry, and those sharing the excitation symmetry.

To further validate this result we compute the spatial profile of the dynamic magnetization from the eigenvector analysis (see methods and Supporting Information S4). We evaluate the mode profile at the fields applied in the experiment, for wavevectors up to 12 rad/μm, reproducing the experimental spatial sampling of 250 nm, and the accessible modes at 9 GHz in both BV and DE configurations.  In addition. to account for the attenuation present in Figure 4b,c, we consider the finite efficiency of the RF transducer.

The corresponding Fourier-transformed signals are shown in Fig. 4d,e and display good agreement with the experimental data in Fig. 4b,c. Notably, the model reproduces the stronger attenuation of the antisymmetric branch without the implementation of any additional symmetry-dependent weighting, suggesting that the observed attenuation arises naturally from the higher density of symmetric modes within the experimentally accessible H-k range, and the spatial localization of the modes (see methods ans Supporting Information). Importantly, the model correctly reproduces the finite separation between the symmetric and antisymmetric branches as well as the associated experimental signal at small, but non-zero, wavevectors.

While our interpretation relies primarily on the analytical framework, full micromagnetic simulations were utilized to rigorously benchmark these results. They confirm that the analytical model accurately reproduces all quantitative dynamical features of the magnetic/nonmagnetic/magnetic stack [70], with $M_s$ = 147 kA/m for both magnetic layers, thicknesses of 95 nm and 93 nm, and anisotropy (other than shape) of 0 and 0.65 kJ $m^{-3}$ for bottom and top magnetic layers respectively, with spacer layer 4 nm, coupled only through the dipolar interaction.

## 4. Conclusion

We have presented a combined experimental and theoretical exploration of spin-wave dynamics in epitaxial magnetic/non-magnetic/magnetic garnet heterostructures. By realizing a vertically coupled trilayer composed of $Y_3Fe_5O_{12}$ / $Y_3Fe_3Al_2O_{12}$ / $Y_3Fe_5O_{12}$ that incorporates a compositionally engineered interlayer with Curie temperature below room temperature, we demonstrate a fully epitaxial platform in which two low-damping magnetic layers interact solely through dynamic dipolar fields. The experimental observations show good quantitative agreement with our general analytical framework for magnetic/non-magnetic/magnetic trilayers, further validated through micromagnetic simulations.

Our results establish robust vertical dipolar coupling in epitaxial garnet multilayers which produces dynamical modes not present in single layer films. In particular, the small difference in anisotropy between the layers breaks the structural symmetry of the heterostructure and provides additional degrees of freedom to tailor the coupled-mode landscape. This anisotropy difference could likely be adjusted or eliminated by making the structure symmetrical, e.g. by including YIAG layers above and below the trilayer. The theoretical model, validated in this work, can be extended to design heterostructures with tunable spin wave dispersion, for example ref. [70] calculates the influence of the composition and thickness of the magnetic thin film and the thickness of the spacer layer on the spin wave dispersion. When combined with lateral confinement, this vertical architecture could be applied to engineer three-dimensional magnonic waveguides, enabling band-structure engineering, control of group velocities, frequency splitting, and anisotropy-dependent non-reciprocal transport.

By uniting ultralow damping, nanoscale epitaxial control, and symmetry-breaking interlayer engineering, vertically coupled garnet heterostructures emerge as a versatile platform for three-dimensional magnonics, with applications in photonic-inspired devices such as magnonic isolators and spin-wave couplers, as well as reconfigurable magnonic circuits, hybrid RF–photonic systems, and low-power wave-based information processing. These results position all-insulating, epitaxial garnet multilayers as a promising foundation for a new generation of spin-based technologies.

**5. Experimental Methods**

## 5.1. Analytical modeling and Micromagnetic Simulations

The analytical model presented in ref. [71] was benchmarked against OOMMF and TetraX simulations. For the OOMMF modeling, we assumed an infinite-film approximation within a computational domain of 200 $\mu m$ × 200 $\mu m$ × 204 nm discretized using a mesh size of 10 nm × 200$\mu m$ × 4 nm. To suppress artificial spin-wave reflections at the sample boundaries, the damping parameter is gradually increased toward the edges using a parabolic profile reaching $\alpha = 1$. The simulations were performed under the Nyquist sampling criterion, ensuring accurate reconstruction of the spin dynamics in Fourier space. A broadband excitation field

$$\overrightarrow{h_{RF}} = (h_{RF_x}, h_{RF_y}, h_{RF_z}) sinc[\omega(t - t_0)]$$

was applied along specific directions depending on the spin-wave configuration. The total simulation time is 20 ns with frequency $\omega = 2\pi \cdot 20$ GHz, which eventually results, because of the Nyquist criterion, in $N = 800$ time steps sampled uniformly every $\Delta t = 25$ ps, resulting then in a frequency resolution of $\Delta f =$ 50 MHz. The magnetization vector field $\vec{m}(x, y, z)$ was collected at each timestep and in each mesh cell, to enable 2D space-time Fourier transformation. A comparative study was then conducted using TetraX simulations based on a multilayer embedded geometry; both approaches were utilized to validate the analytical model. From this preliminary characterization, we extrapolated the equivalent out-of-plane uniaxial anisotropy $K_u = 0.65$ kJ/$m^3$, and refined the computational $M_s$ to $1.47 \cdot 10^5$ A/m. Using this refined model, it was then possible to directly access the eigenvectors of the structure and simulate the expected phase-resolved Brillouin light scattering (PR-BLS) signal.

We retrieved the eigenvalues and their associated eigenvectors for each experimental magnetic field. By inspecting the available modes at the experimental frequency of 9 GHz, we constructed a spatial matrix comprising the magnetization mode shapes at each $\omega$ and k for the available modes and reconstruct the spatial profile (see Supporting Information S3)

$$m(x, \omega) = \sum \frac{|m^p(k_i)|}{|m^p(k_i)| + |m^q(k_i)|} \exp(j[k_i x + \angle m^p(k_i)]),$$

Where i= symmetric, antisymmetric and p=1,2 the layer index (i.e. top or bottom). Finally, we computed the expected efficiency of the RF transducer in k-space [80] and multiplied it by the fast Fourier transform (FFT) of the magnetization. This procedure replicated the exact spatial sampling used in the experiment, ultimately yielding the power spectral density of the signal.

## 5.2. Sample and antenna fabrication

Prior to the film deposition, the as-received (111)-oriented $Gd_3Ga_5O_{12}$ substrates were annealed in a furnace tube at 1000 °C for 6 hours in oxygen to smooth the surface. The $Y_5Fe_3O_{12}$ (100 nm)/$Y_5Fe_3Al_2O_{12}$(4 nm)/$Y_5Fe_3O_{12}$(100 nm)/$Gd_3Ga_5O_{12}$ (111) heterostructure was grown by PLD (Neocera) with a Coherent KrF excimer laser (wavelength of 248 nm). The laser energy was set between 300-600 mJ to reach a fluence of 2 J $cm^{-2}$ at the target surface with a pulse frequency of 10 Hz. Prior to the film growth, a base pressure of 5 x $10^{-6}$ Torr was reached, and then 150 mTorr of oxygen was maintained while the substrate was heated to 650 °C at a ramp rate of 20 °C $min^{-1}$. All three layers were grown sequentially, without changing pressure or temperature in the chamber. The YIG layers were grown from a stoichiometric YIG target, whereas, the nonmagnetic layer, $Y_5Fe_3Al_2O_{12}$, was grown by co-deposition between the same YIG target and a stoichiometric YAG target. The targets were alternatively ablated with a pulse rate of 10 Hz and a ratio 3:1 YIG:YAG as the YAG target had a higher deposition rate than that of the YIG target. Following the deposition, the sample was cooled to room temperature at the same ramp rate. Radio frequency transducers made by Ti(10 nm)/ Au(100 nm) were then deposited by lift off using electron beam lithography. We adopt a strip line design with short dimension of 80, 160 nm width and 100 $\mu m$ length, in a ground–signal–ground (GSG) electrical scheme (see Supporting Information S2).

### 5.3. Structural and Magnetic Characterization

Magnetic hysteresis loops were measured on a Vibrating Sample Magnetometer (Digital Measurement Systems Model 1660), with a field resolution of 0.05 T, maximum field of 1 T, and the field applied in the plane of the substrate. After the measurement, a linear paramagnetic substrate contribution was subtracted. HRXRD was performed on a Rigaku Smartlab multipurpose diffractometer with a Cu Kα X-ray source. Rigaku Globalfit integrated thin film analysis software was used to analyze diffraction data by first fitting to data from single layer films then combining the fit results to the tri-layer heterostructure data. A Thermo Fisher Talos F200i was used to acquire all HAADF-STEM and EDX data, and the Velox software was used to process the images and color maps.

### 5.4. FMR

The sample was mounted face down on a dedicated printed circuit board (PCB) featuring an integrated GSG coplanar waveguide (33 $\mu m$ trace width). A single-tone radio frequency (RF) signal was applied, sweeping from 1 to 16 GHz, while the external magnetic field was varied with a resolution of 0.1 mT. The setup consists of a double-modulation within a double lock-in detection scheme. This technique enables high-sensitivity detection of weak magnetic resonance signals by combining frequency and magnetic field modulation with synchronous demodulation. The radiofrequency excitation was provided by an RF generator modulated with a square wave at 30 kHz and injected into the dedicated FMR PCB. To enable phase-sensitive detection, a small sinusoidal magnetic field modulation was superimposed on the static bias field using modulation coils. The magnetic field was modulated at a frequency of 270 Hz with an amplitude of 5 mT. The transmitted microwave signal was detected via an RF diode detector, which amplified and converted the RF power into a voltage with a gain of 20 dB. This voltage was subsequently demodulated in two stages using lock-in amplifiers, yielding the mixed derivative of the signal for both frequency and magnetic field. This dual-demodulation approach significantly enhances the signal-to-noise performance, suppresses baseline drifts, and improves sensitivity to small magnetic absorption features.

### 5.5. Micro-focused Brillouin light scattering

We used a custom-developed optical setup to measure Brillouin light scattering spectra [79], [81], [82]. A single-mode laser (COBOLT Samba) with a wavelength of 532 nm was used as light source. The spectral purity of the laser light was improved by a Fabry-Perot filter (TCF-2, Table Stable). An optical microscope with active stabilization was used to compensate the mechanical drifts of the sample (THATec Innovation). The light was focused and collected through the same objective lens (Zeiss LD EC Epiplan-Neofluar 100 × /0.75 BD). The inelastic frequency shift was measured with a tandem Fabry-Perot interferometer (TFP-2HC interferometer, table stable) [83]. To generate the magnetic field, we used a water-cooled GMW 5403 electromagnet powered by two KEPCO BOP20-20DL power supplies and a predefined current field calibration at the sample position.

**Acknowledgments**

The authors acknowledge valuable discussions with R. Bertacco and F. Maspero. L. Menna acknowledges G. Gubbiotti and M. Madami for valuable discussions. S. Kurdi acknowledges support from Dutch Research Council (NWO) via VI.Veni.222.296. O. Wojewoda acknowledges support from a Marie Curie Fellowship (Horizon Europe - MSCA grant agreement No. 101211677, project FeriMag ). C. A. Ross and M. J. Gross acknowledge support from NSF DMR 2323132. A. Del Giacco acknowledges support from Politecnico di Milano. Shared facilities of MIT.nano were used in this work. The authors would also like to thank David Bono for his help with the experimental setup of the FMR station, and Aubrey N. Penn for her help with the TEM measurements.

**Data Availability Statement**

Data will be provided by the authors on reasonable request.

# Supporting Information

**Contents**

## S1: Composition of Al-substituted YIG Interlayer

A molecular field coefficient model [1,2] was used to determine the variation of the Curie temperature of $Y_3Fe_{(5-x)}Al_xO_{12}$ with composition *x*. Al has a tetrahedral site preference but occupies both sites, so the model assumes 60% of the Al on the tetrahedral site and 40% on the octahedral sites. Lattice parameter was assumed to vary linearly with Al content according to Vegard's law. The composition with room temperature paramagnetism and the largest lattice parameter (i.e. lowest *x*) was selected to minimize the strain in the interlayer grown on YIG.

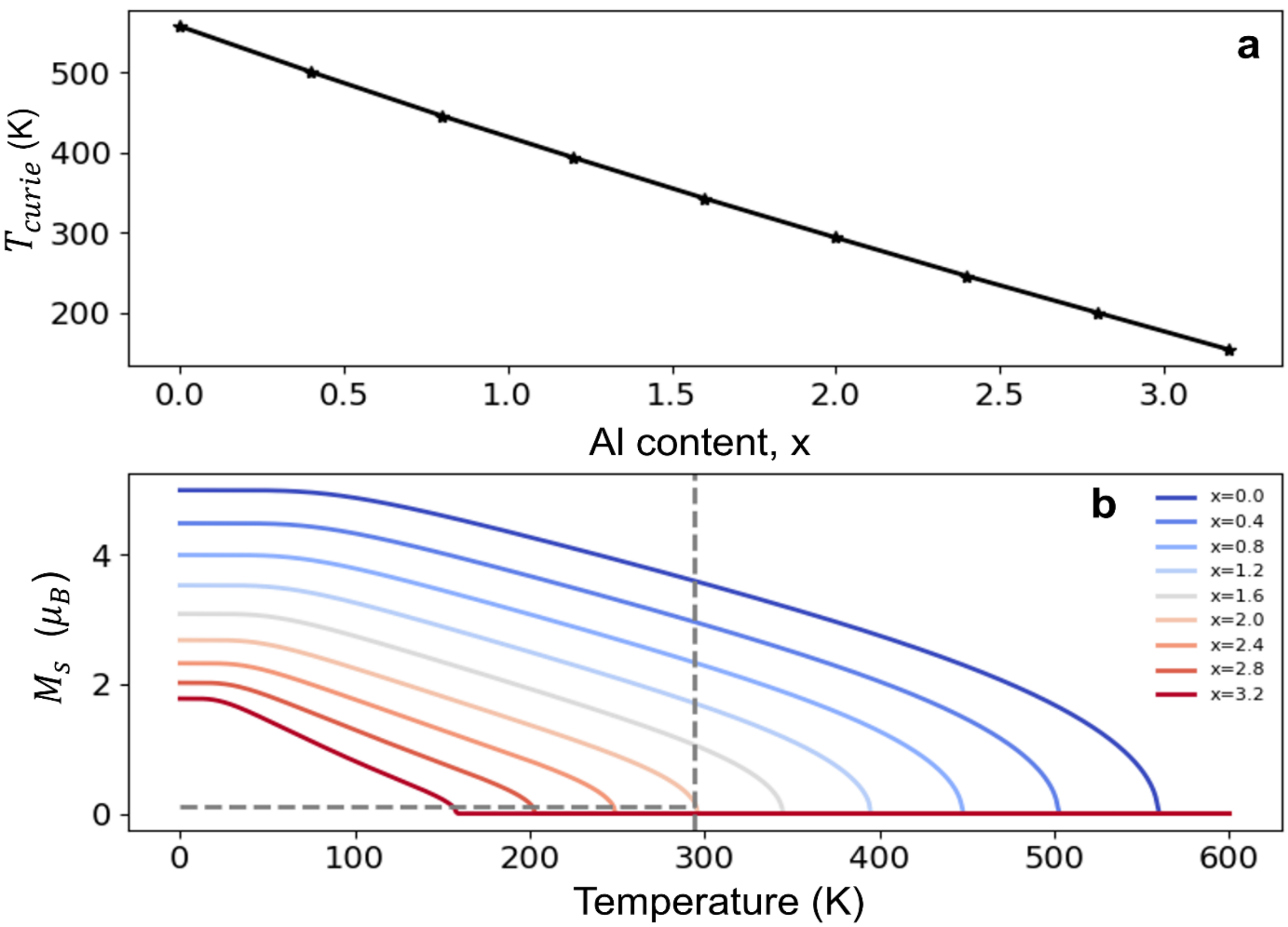

**Figure S1 (a)** Curie temperature plotted as a function of Al substitution into YIG. Increasing the Al concentration reduces the Curie temperature of the garnet and the lattice parameter. **(b)** Moment per formula unit versus temperature of Al doped YIG with various levels of doping. Al is assumed to preferentially occupy the tetrahedral d-site due to its smaller size (60% tetrahedral, 40% octahedral), which lowers the moment across all temperatures. A composition of $Y_3Fe_3Al_2O_{12}$ yields a Curie temperature of 294 K according to the model.

**S2: Dynamics Localization**

To validate the spatial localization of the modes presented in the main text (Fig. 2) and the attribution of the FMR peaks to the top or bottom layer, TetraX simulations are reported in which the localization behavior at k = 0 and for the Damon–Eshbach modes is clearly visible. The simulations were carried out within a multilayer embedded geometry, weighting the localization of each (f, k) point on the spatial mesh and attributing a colormap based on the localization between the two layers. The simulations were performed for two YIG layers with thicknesses of 95 and 93 nm, with saturation magnetization $\boldsymbol{M_s}$ = 147 kA/m, exchange stiffness A = 4 pJ/m, and uniaxial out of plane anisotropy $\boldsymbol{K_u}$ = 0.65 kJ/m$^3$ only in the top layer for an external applied field $\boldsymbol{H_a}$=245 mT.

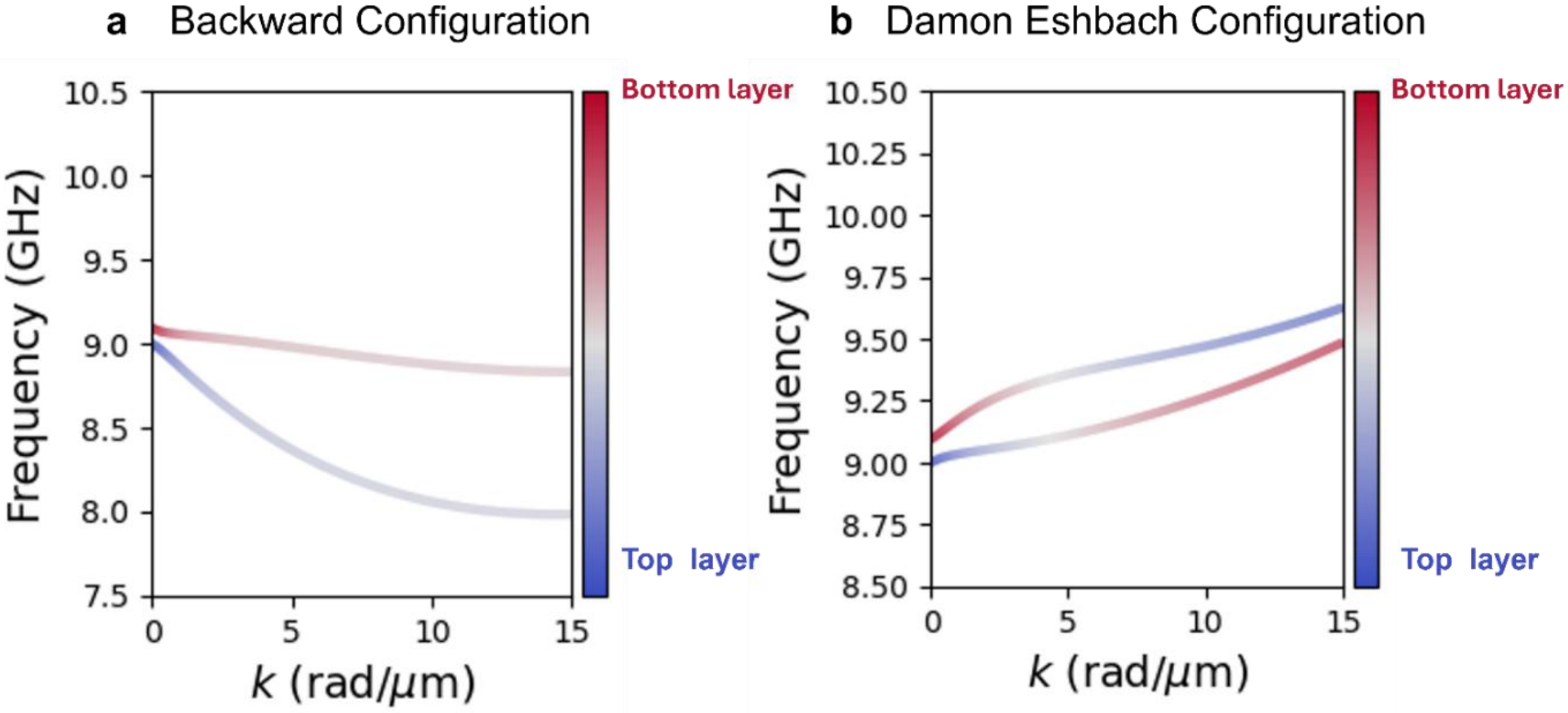

**Figure S2. (a)** Backward (BV) and **(b)** Damon–Eshbach (DE) dispersion relations. The colormap encodes the spatial localization of each mode, with red indicating predominant localization in the bottom layer (higher frequency) and blue in the top layer (lower frequency). The well-defined localization at *k* = 0 for both configurations, and the delocalized and localized characters of the BV and DE modes, respectively, are consistent with the main text (Fig. 2b,c).

**S3: Excitation efficiency**

To characterize the fabricated antenna and its excitation selectivity, we measure the geometry (Fig. S2 a) and evaluate the excitation efficiency as a function of the wavevector. The efficiency provides a direct estimate of the $k$-space bandwidth of the lithographically defined electric transducer (Fig. S2 b). To further investigate the relative excitation efficiency of the symmetric and antisymmetric modes within the experimental field range and frequency, we simulated the available modes for both the DE and BV configurations (Fig. S3a, b). The results reveal a favored symmetric state which, in conjunction with the symmetric torque, explains the weakness of the antisymmetric signal reported in the main text.

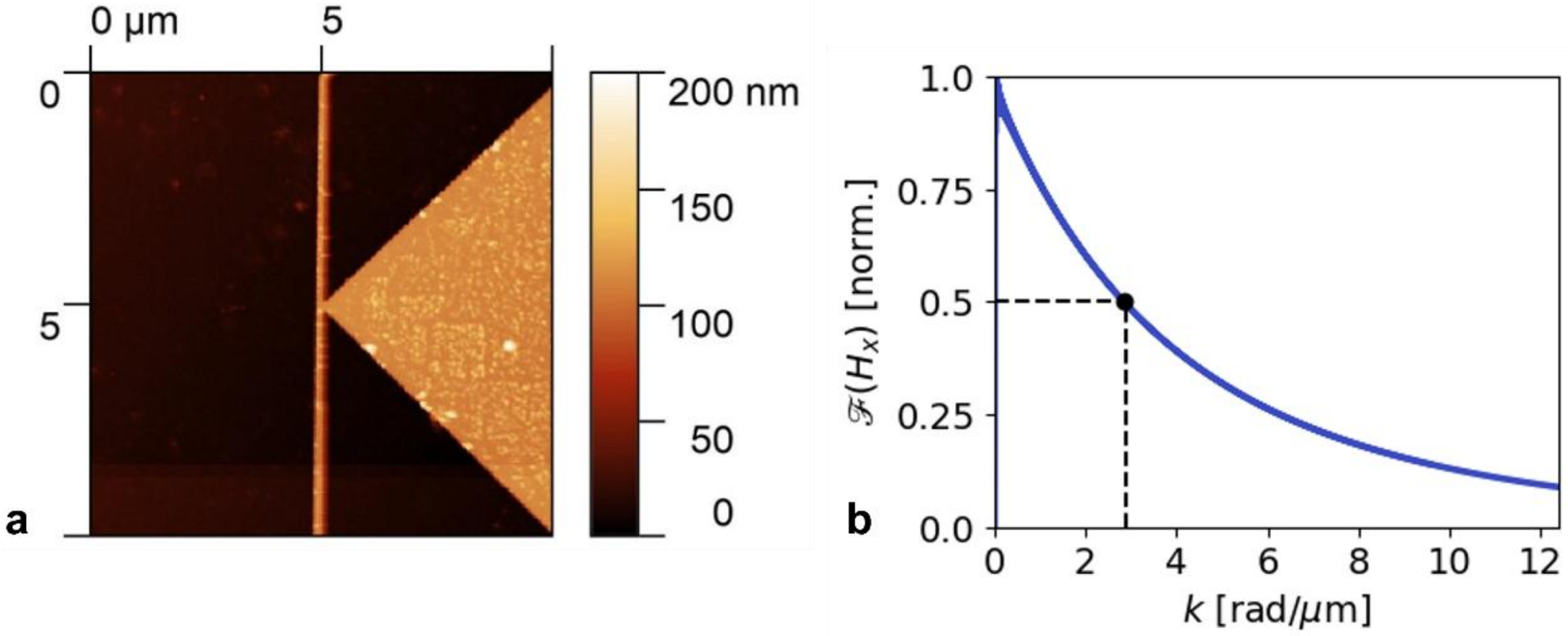

**Figure S3.1 (a)** AFM image of one of the fabricated electric transducers. **(b)** Excitation efficiency as a function of the wavevector $k$. Black markers and dashed lines indicate the wavevector at which the efficiency drops to half of its maximum value ($\approx 2.8\, rad/\mu m$). The efficiency is extracted from the Fourier amplitude of the excited mode and reflects the spatial selectivity of the lithographically defined antenna. This characteristic wavevector defines the effective $k$-space bandwidth of the transducer.

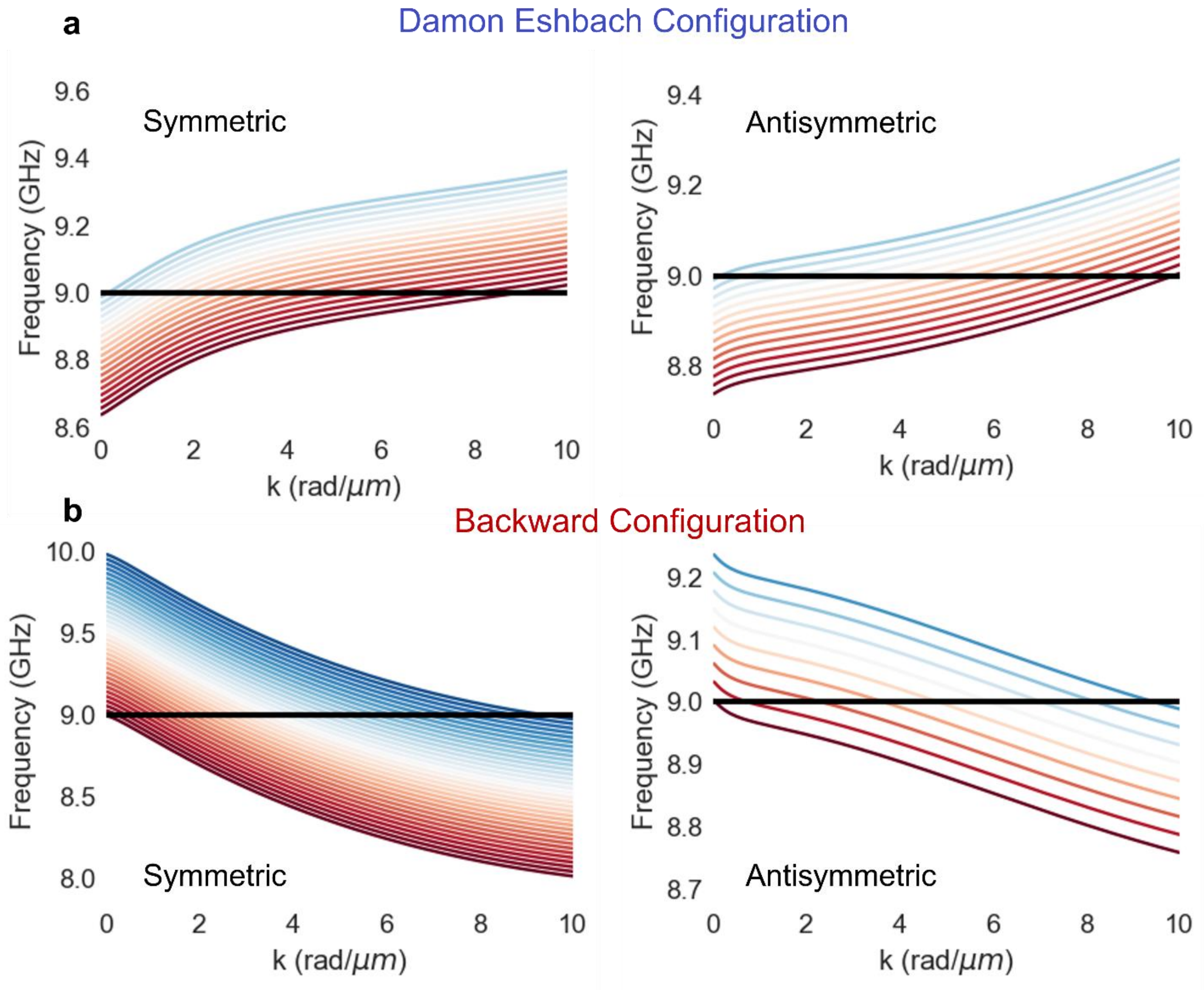

**Figure S3.2**. **(a)** Dispersion relation for DE configuration highlighting the available modes at the experimental driving frequency (black line) and swept fields, for the symmetric and antisymmetric branch. **(b)** Dispersion relation for BV configuration highlighting the available modes at the experimental driving frequency (black line) and swept field, for the symmetric and antisymmetric branch.

## S4. Brillouin light scattering profiles

To further validate the model and simulations used to interpret the experimental results, we compare the spatial distribution of the excited spin waves measured by Brillouin light scattering (BLS) with the corresponding micromagnetic simulations. The measurements are performed at a fixed excitation frequency (Fig. S4). In Figure S5 we report the overlap between the experimental data and the modeling presented in the main text.

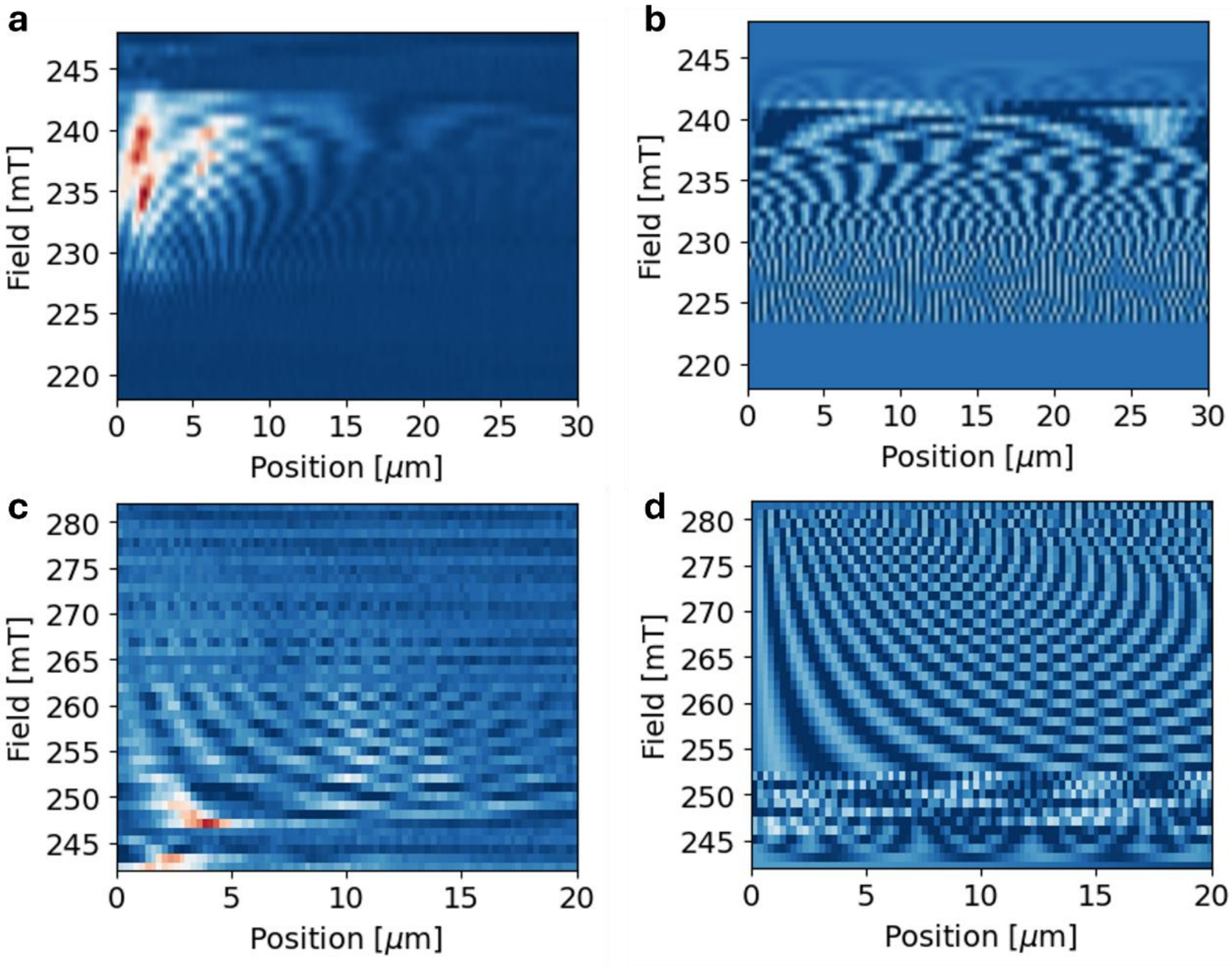

**Figure S4.** Experimental **(a,c)** and simulated **(b,d)** BLS spatial scans at a constant excitation frequency $f = 9\ GHz$ for the Damon–Eshbach (DE) and backward volume (BV) configurations, respectively. The maps show the spatial distribution of the spin-wave intensity excited by the antenna.